\documentclass[reprint,pra,notitlepage,floatfix]{revtex4-1}
\usepackage{newtxtext}
\usepackage{newtxmath}
\let\lambda\lambdaup

\usepackage{amsmath}
\usepackage{amssymb}
\usepackage{graphicx}
\usepackage{color}
\usepackage[colorlinks=true,linkcolor=blue,citecolor=blue,urlcolor=blue]{hyperref}
\graphicspath{{fig/}}

\allowdisplaybreaks
\begin{document}

\title{Mixed spectra and partially extended states in a two-dimensional quasiperiodic model}
\author{Attila Szab\'o}
\author{Ulrich Schneider}
\affiliation{Cavendish Laboratory, University of Cambridge, Cambridge CB3 0HE, United Kingdom}

\begin{abstract}
    We introduce a two-dimensional generalisation of the quasiperiodic Aubry--Andr\'e model. 
    Even though this model exhibits the same duality relation as the one-dimensional version, its localisation properties are found to be substantially more complex.
    In particular, partially extended single-particle states appear for arbitrarily strong quasiperiodic modulation. They are concentrated on a network of low-disorder lattice lines, while the rest of the lattice hosts localised states.
    This spatial separation protects the localised states from delocalisation, so no mobility edge emerges in the spectrum. Instead, localised and partially extended states are interspersed, giving rise to an unusual type of mixed spectrum and enabling 
    complex dynamics even in the absence of interactions. A striking example is ballistic transport  across the low-disorder lines while the rest of the system  remains localised.
    This behaviour is robust against disorder and other weak perturbations. 
    Our model is thus directly amenable to experimental studies and promises fascinating many-body localisation properties.
\end{abstract}

\maketitle

\section{Introduction} 

Quasiperiodic systems have sparked interest among physicists since the discovery of the Hofstadter butterfly~\cite{Harper1955SingleField,Azbel1964EnergyField,Hofstadter1976EnergyFields} and of quasicrystalline materials which combine long-range positional order  with crystallographically forbidden rotational symmetries~\cite{Shechtman1985TheAl6Mn}. More recently, quasiperiodic models have received renewed attention in cold atom experiments on disordered quantum gases, Bose glasses, and many-body localisation (MBL)~\cite{Roati2008AndersonCondensate,Deissler2010DelocalizationInteractions,Sanchez-Palencia2010DisorderedControl,Schreiber2015ObservationLattice,Bordia2016CouplingSystems,Bordia2017ProbingSystems,Luschen2017ObservationSystems,Viebahn2019Matter-WaveLattice,Corcovilos2019Two-dimensionalExperiments,Kohlert2019ObservationEdge}. 

Such models present a wide range of intriguing localisation properties without match among randomly disordered systems, including critical spectra and multifractal eigenstates away from phase transitions~\cite{Kohmoto1983LocalizationEscape,Ostlund1983One-DimensionalPotential,Kohmoto1986QuasiperiodicDiffusion,You1991GeneralizedWavefunctions,Han1994CriticalCoupling,Liu2015LocalizationModel} and Anderson localisation transitions in one dimension~\cite{Aubry1980AnalyticityLattices,Han1994CriticalCoupling,Liu2015LocalizationModel}, as well as unusual transport properties in higher dimensions~\cite{Devakul2017AndersonPotentials,Sutradhar2019TransportSystems}.
Furthermore, quasiperiodic potentials contain no rare regions in the usual sense, i.e., patches in which the local disorder is by chance substantially lower or higher than on average. Such regions give rise to Griffiths effects~\cite{Vojta2010QuantumExperiment} which are expected to substantially affect MBL in disordered one-dimensional systems~\cite{Gopalakrishnan2016GriffithsSystems,Khemani2017TwoTransition} and might destabilise MBL completely in higher dimensions~\cite{DeRoeck2017StabilitySystems}. Quasiperiodicity might therefore prove essential to stabilising MBL in higher dimensions.

Quasiperiodic systems also inherit fascinating topological properties from higher-dimensional periodic parent Hamiltonians~\cite{Kraus2012TopologicalQuasicrystals} from which they can be derived using cut-and-project methods~\cite{Senechal1995QuasicrystalsGeometry}. As an example, two-dimensional quasicrystals can exhibit the four-dimensional integer quantum Hall effect~\cite{Kraus2013Four-DimensionalQuasicrystal,Zilberberg2018PhotonicPhysics}.

\begin{figure}
    \centering
    \includegraphics{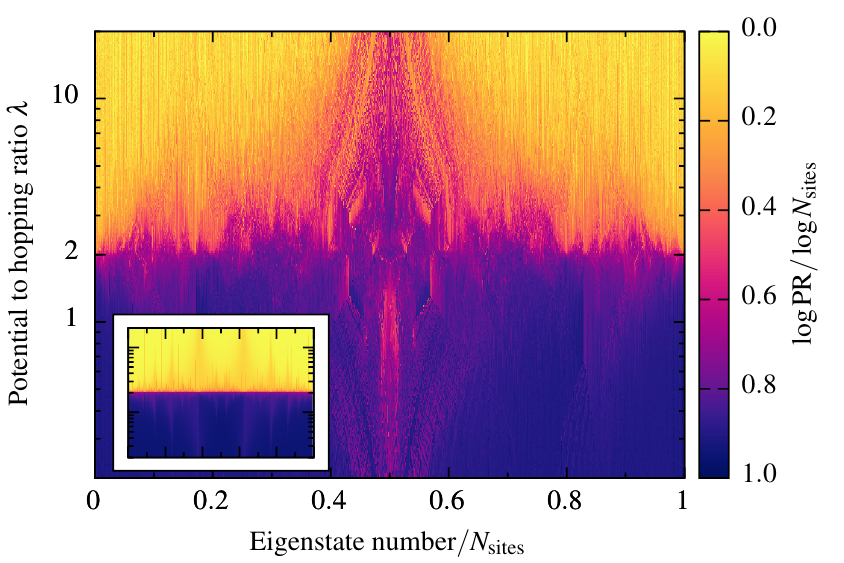}
    \caption{Participation ratios of all eigenstates of the 2DAA model as a function of $\lambda$ for $\beta=70/99$ ($N_\mathrm{sites}=9801$). A large number of partially extended states persist at $\lambda\gg2$ in the middle of the spectrum; due to Aubry duality, a similar number of eigenstates is not fully extended for $\lambda\ll2$. Inset: participation ratios for the 1DAA model with $\beta=987/1597$ ($N_\mathrm{sites}=1597$). A sharp localisation transition occurs at the self-dual point $\lambda=2$.}
    \label{fig: spectra}
\end{figure}

In this paper, we consider a two-dimensional generalisation of the celebrated Aubry--Andr\'e (1DAA) model~\cite{Aubry1980AnalyticityLattices},
\begin{align}
    H &= -J \sum_n \big( a^\dagger_{n}a^{\phantom\dagger}_{n+1} + \mathrm{H.c.}\big) - \lambda J\sum_n \cos(2\pi\beta n) a^\dagger_{n}a^{\phantom\dagger}_{n},
    \label{eq: 1d AA model}
\end{align}
which we shall call the two-dimensional Aubry--Andr\'e (2DAA) model. It is defined on a square lattice by the Hamiltonian
\begin{align}
    H &= -J \sum_{nm} \big[a^\dagger_{nm} \big(a^{\phantom\dagger}_{n+1,m} + a^{\phantom\dagger}_{n,m+1}\big) + \mathrm{H.c.}\big] 
    \label{eq: 2d AA model}\\
    &- \lambda J\sum_{nm} \big\{\cos[2\pi\beta(n+m)]+\cos[2\pi\beta(n-m)]\big\}a^\dagger_{nm}a^{\phantom\dagger}_{nm}.\nonumber
\end{align} 
In both models, $\beta\not\in\mathbb{Q}$ and $\lambda$ are the incommensurate wave number and dimensionless amplitude of the quasiperiodic modulation, and the $a^\dagger$ are creation operators living on the lattice sites. 
This form of quasiperiodic modulation can readily be incorporated into existing optical lattice experiments using two additional weak one-dimensional (1D) lattices at $45^\circ$ to the main lattice axes.
Throughout the paper, we set $\beta=1/\sqrt2$, which amounts to setting the wave vectors of the principal lattice and the perturbation equal~\cite{Viebahn2019Matter-WaveLattice}. We emphasise that this model is non-separable and hence fundamentally different from earlier separable ones where the non-interacting localisation transition was directly controlled by the underlying 1D Hamiltonians~\cite{Bordia2016CouplingSystems,Bordia2017ProbingSystems,Rossignolo2019LocalizationChains}.

Both~\eqref{eq: 1d AA model} and~\eqref{eq: 2d AA model} admit an Aubry duality transformation~\cite{Aubry1980AnalyticityLattices,Devakul2017AndersonPotentials}; that is, they can be reexpressed in momentum space in the same form, with the parameter $\lambda$ changed to $4/\lambda$ (see Appendix~\ref{app: aubryduality} for details of the transformation). 
In the 1DAA model, this induces a localisation transition at the self-dual point $\lambda=2$: All eigenstates are exponentially localised for $\lambda>2$ and extended for $\lambda<2$~\cite{Aubry1980AnalyticityLattices,Suslov1982LocalizationSystems,Soukoulis1982LocalizationPotentials,Jitomirskaya1999Metal-InsulatorOperator} (see inset of Fig.~\ref{fig: spectra}).

We find in this paper that this is not the case for the 2DAA model: Namely, some states remain partially extended even for $\lambda\gg2$, as illustrated in Fig.~\ref{fig: spectra}. Furthermore, these states are not separated from localised ones by a mobility edge, as expected on general grounds \cite{Mott1987The1967}, but localised and partially extended states are interspersed in the spectrum. 
We explain this behaviour in terms of weakly disordered lines that appear deterministically in the quasiperiodic potential~\eqref{eq: 2d AA model}; these also lead to strongly inhomogeneous expansion dynamics that could be detected experimentally. 
%
%

\begin{figure}
    \centering
    \includegraphics{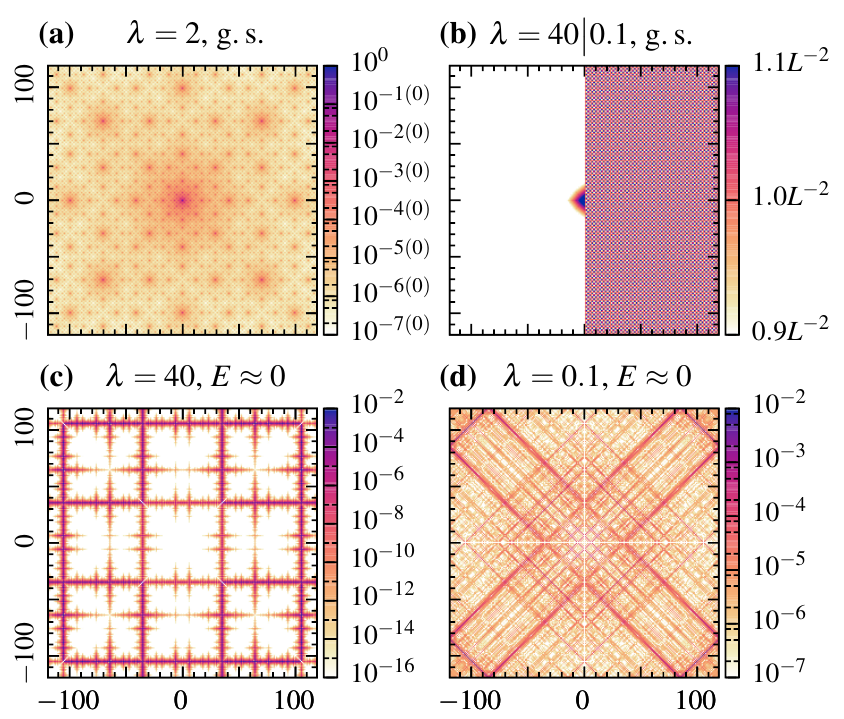}
    \caption{Wave function weight $|\psi_{nm}|^2$ for selected eigenstates of the 2DAA model with $\beta=169/239$ ($L=239$). 
    \textbf{(a)} Ground state at the self-dual point $\lambda=2$ (log scale from $10^{-7}$ to $10^0$), which shows a multifractal structure similar to the critical eigenstates of the 1DAA model.
    \textbf{(b)} Ground state at $\lambda=40$ (left half, log scale from $10^{-70}$ to $10^0$) and at $\lambda=0.1$ (linear scale to the right). The former is strongly localised with a very small localisation length ($\xi\approx0.1$); the latter is fully extended.
    \textbf{(c)} Median energy state ($E\approx 0$) at $\lambda=40$. The wave function is concentrated on a few lattice lines, along which the quasiperiodic potential is weaker than on average.
    \textbf{(d)} Median energy state ($E\approx0$) at $\lambda=0.1$. The wave function is concentrated on a few diagonal lines of the lattice that arise as Aubry duals of the lines in (c).
    }
    \label{fig: wave function}
\end{figure}

\section{Partially extended states at strong disorder}
\label{sec: partially extended}
We used exact diagonalisation to obtain the full single-particle spectrum of the 2DAA model. We replaced the irrational $\beta=1/\sqrt{2}$ with close rational approximations $\beta = M/L$ derived from its continued fraction expansion~\cite{Lang1995IntroductionApproximations}: This allows us to use periodic boundary conditions on a square lattice of $N_\mathrm{sites} = L\times L$ sites. For each normalised eigenstate $|\psi\rangle$, we evaluated its participation ratio (PR), defined as
\begin{equation}
    \mathrm{PR} = \Big(\sum_{nm} \big|\psi_{nm} \big|^4\Big)^{-1};
    \label{eq: PR wave function}
\end{equation}
for a wave function evenly distributed on $k$ sites, $\mathrm{PR}=k$. We plot PR as a function of $\lambda$ and position in the spectrum in Fig.~\ref{fig: spectra}, together with the equivalent results for the 1DAA model (inset). In the latter, one can clearly see a localisation transition in all eigenstates at $\lambda=2$, with PR close to either 1 or the number of sites everywhere except for a narrow region at $\lambda\approx2$. The phase diagram of the 2D model is much more complex: There is no sharp transition at the self-dual point, but localised and delocalised states coexist in a wide region around it. Most notably, there is a ``funnel'' of eigenstates that appear neither fully localised nor fully extended ($\mathrm{PR}\sim N_\mathrm{sites}^{0.5}$) in the middle of the spectrum, interspersed with either localised (for $\lambda\gg 2$) or extended (for  $\lambda\ll2$) states. 

To further illustrate these features, we plot several representative eigenstates of the 2DAA model in Fig.~\ref{fig: wave function}. The ground states (top panels) follow a similar pattern as in the 1DAA model: They show fractal properties at the self-dual point and are extended and exponentially localised on either side of it.  In the middle of the spectrum, however, the picture away from $\lambda=2$ is very different (bottom panels). For $\lambda\gg2$, we find many states where most of the wave function weight is concentrated on a small number of horizontal and vertical lines, with small, exponentially decaying weight close to them. Indeed, all eigenstates with significant PR follow this pattern and populate the same set of lines. For $\lambda\ll2$, we see a similar, although less sharp, pattern concentrated on a few diagonal lines. We note that these two types of wave functions transform into each other under the Aubry duality transformation that includes both a Fourier transform and a $45^\circ$ rotation (see Appendix~\ref{app: aubryduality}).

\begin{table}
    \centering
    \begin{tabular}{|c|c|c||c|c|c|c|c|c|} \hline
        $n$ &
        $\pm35$ & $\pm105$ & $\pm64$ & $\pm6$ & $\pm76$ & $\pm93$ & $\pm23$ & $\pm47$\\ \hline
        $\left|\tilde\lambda_n\right|$ & 
        0.526 &  1.577 &  2.628 &  3.679 &  4.729 & 5.779 & 6.827 & 7.874 \\\hline
    \end{tabular}
    \caption{The eight smallest local disorder amplitudes $\tilde\lambda$ for the 2DAA model with $\beta=169/239$ and $\lambda=40$. $|\tilde\lambda|<2$ for the first two; the resulting extended 1DAA eigenstates account for the bulk of the statistical weight in Fig.~\ref{fig: wave function}(c).
    Subsequent lines define localised 1DAA models: these appear in the plotted state as progressively shorter barbs close to the extended lines.}
    \label{tab: lambda}
\end{table}

The origin of these eigenstates at large $\lambda$ can be explained by rewriting the potential term of \eqref{eq: 2d AA model} as
\begin{align}
    V_{nm} &= \lambda J \big\{\cos[2\pi\beta(n+m)]+\cos[2\pi\beta(n-m)]\big\}\nonumber\\
    &= 2\lambda J \cos(2\pi\beta n)\cos(2\pi\beta m) = \tilde{\lambda}_n J \cos(2\pi\beta m).
\end{align}
Since the effective disorder amplitudes $\tilde\lambda_n =2\lambda\cos(2\pi\beta n)$ along lattice lines form a quasiperiodic sequence, there will always exist lines for which $|\tilde\lambda|\ll\lambda$, i.e., lines along which the disorder is much weaker than it typically is across the system. If one removed the horizontal hopping terms from \eqref{eq: 2d AA model}, the remaining model would consist of independent 1DAA Hamiltonians with parameter $\tilde\lambda_n$: along lines where $|\tilde\lambda_n|<2$, all eigenstates  would be extended~\cite{Aubry1980AnalyticityLattices}. 
Reintroducing the horizontal hopping terms will then hybridise the 1DAA eigenstates on different lines. By the same argument, horizontal lines also form 1DAA models, some of which are in the extended phase: these hybridise with the aforementioned vertically extended states, leading to mesh-like eigenstates living on the quasiperiodic grid of low-disorder lines, as seen in Fig.~\ref{fig: wave function}. 
Since a finite fraction of approximately $2/(\pi\lambda)$ of all lines is extended, the PR of these states will, for sufficiently large system sizes, scale as $L^2$, i.e., extensively. On intermediate scales, however, they appear one-dimensional, with their two-dimensional character limited to the intersections of horizontally and vertically extended lines: therefore, we call these states \textit{partially extended}.

\begin{figure}
    \centering
    \includegraphics{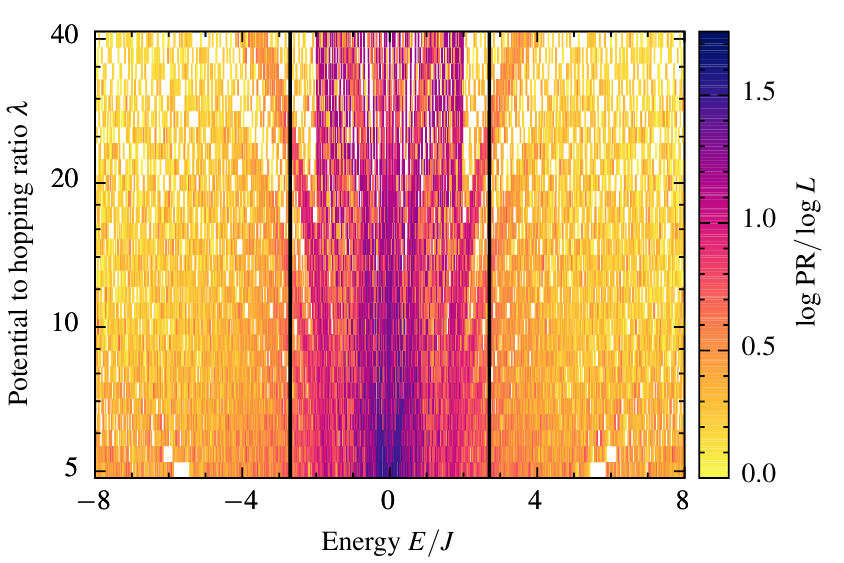}
    \caption{Participation ratios of the eigenstates of the 2DAA model close to zero energy, as a function of $\lambda$ and energy, for $\beta=169/239$ ($L=239$). Regardless of the value of $\lambda$, a large number of eigenstates with energies within the bandwidth~\cite{Note21} of the critical ($\lambda=2$) 1DAA model (black lines) are partially extended and have large ($L^1..L^{1.7}$) PR; no such states occur outside of this energy window.}
    \label{fig: energy dependence}
\end{figure}

On the other hand, the exponential localisation of 1DAA lines with $|\tilde\lambda|>2$ suppresses the statistical weight of partially extended states away from the low-disorder lines.
As a result, partially extended and localised states at similar energies do not hybridise. This prevents the formation of clean mobility edges~\cite{Mott1987The1967} and instead gives rise to the observed mixed spectra, similar to other quasiperiodic systems~\cite{Guo2014DelocalizationSystems,Li2014RealizationTransparency,Chandran2017LocalizationGlass,Huang2019MoireSystems,Kariyado2019FlatLattices}.

\begin{figure}[!h]
    \centering
    \includegraphics{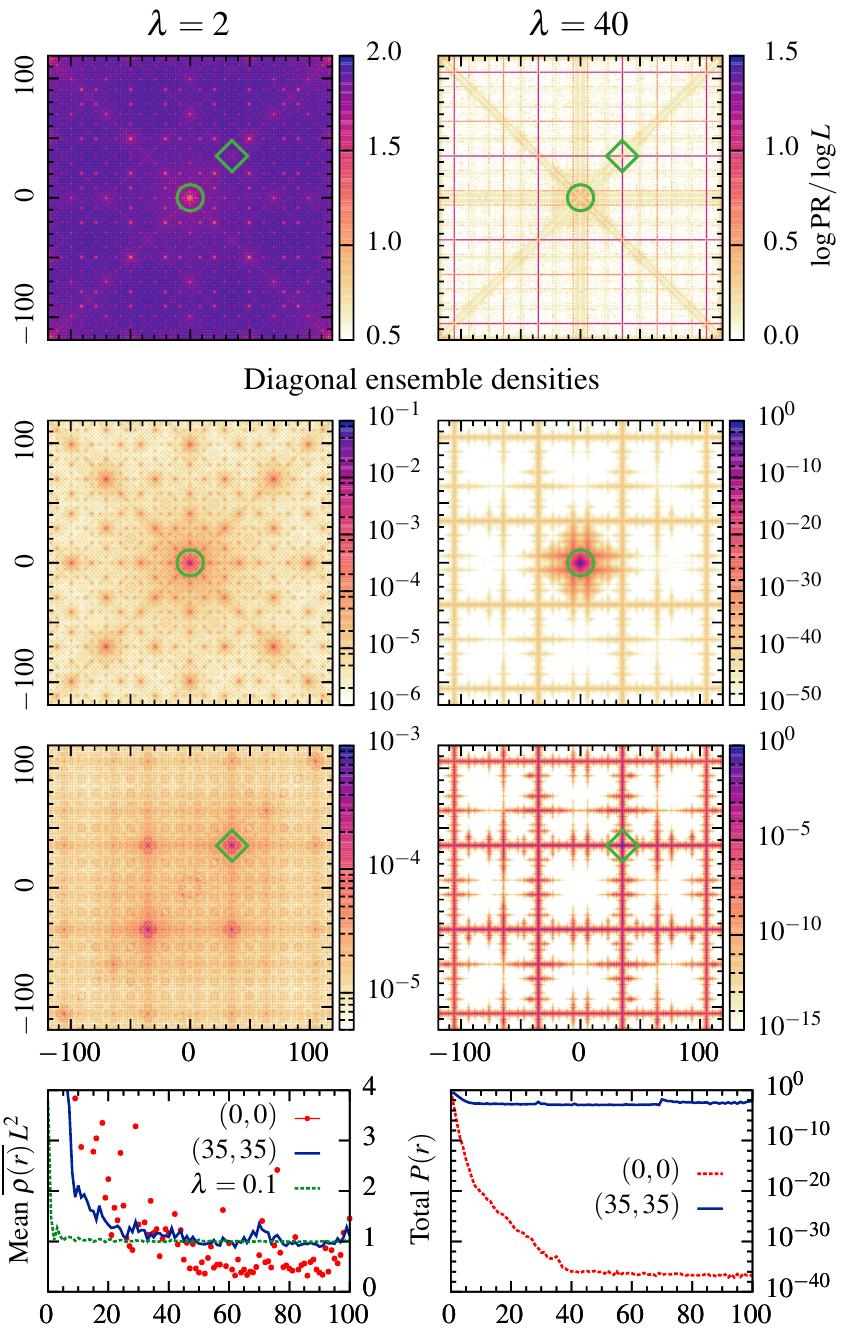}
    \caption{\textit{Top panels:} Diagonal ensemble PR~\eqref{eq: diagonal ensemble PR} as a function of the initial site for $\lambda=2$ and 40. At the self-dual point, the distribution appears almost uniformly delocalised for most initial sites; at $\lambda\gg2$, by contrast, all sites except those along low-disorder lines show strongly localised dynamics. 
    \textit{Middle panels:} Diagonal ensemble densities $\varrho_{nm}(n',m')$ for initial sites $(n,m) = (0,0)$ and $(35,35)$ (green symbols) for $\lambda=2,40$. At the self-dual point, special initial sites like the origin lead to pronounced fractal dynamics; for more generic sites, however, the late-time distribution is almost uniform. At strong modulation, delocalised dynamics occurs only on low-disorder lines.
    \textit{Bottom panels:} 
    The mean radial density $\rho(r)$ tends to a constant $\approx L^{-2}$ for almost all initial sites for $\lambda\le 2$, indicating 2D extended dynamics [left; blue and red: $\lambda=2$; green: $\lambda=0.1$ (35,35)]. 
    For $\lambda\gg2$ (right), initial sites on a low-disorder line lead to largely constant total radial probability $P(r)$, indicating partially extended (quasi-1D) dynamics (blue curve); for a generic initial site (red curve), exponential decay is capped at very low levels by the contribution of delocalised lines. }
    \label{fig: dynamical participation ratio}
\end{figure}

It is indeed easy to verify that the state shown in Fig.~\ref{fig: wave function}(c) is extended precisely along the lines with $|\tilde\lambda|<2$ and that for $|\tilde\lambda|>2$, the effective localisation length decreases with growing local amplitude (see Table~\ref{tab: lambda}). 
Furthermore, since the partially extended eigenstates are effectively superpositions of extended 1DAA eigenstates, we expect their energies to lie within the spectrum of the relevant 1DAA model.  In particular, they should only occur between the lowest and highest eigenvalues of the critical ($\lambda=2$) 1DAA Hamiltonian, as these bound the spectrum of \eqref{eq: 1d AA model} for all $\lambda<2$~\footnote[21]{For $\beta=1/\sqrt2$, we find these bounds by exact diagonalisation to be $\approx\pm2.703 J$; see black lines in Fig.~\ref{fig: energy dependence}.}.
We show that this is indeed the case in Fig.~\ref{fig: energy dependence}: The PR of eigenstates outside of the critical bandwidth is substantially smaller than the large values attained by many eigenstates inside. Note, however, that some states remain localised even inside this bandwidth: these appear on lattice sites with small $|V_{nm}|$ away from the extended lines. This gives rise to a peculiar mixed spectrum with interspersed localised and partially extended states.
%
%

\section{Dynamics} 
\label{sec: dynamics}
In experimental settings, dynamics following a quantum quench from a given initial state is often easier to access than individual eigenstates. The coexistence of localised and partially extended eigenstates at $\lambda\gg2$ and the fact that the latter are largely confined on a special set of low-disorder lattice lines indicate that the same lines will also show peculiar dynamical properties. To confirm this intuition, we have considered the expansion dynamics of states initially localised on a single lattice site $(n,m)$ under the 2DAA Hamiltonian. In particular, we focus on the long-time average of density, given by the diagonal ensemble
\begin{equation}
    \rho_{nm} = \sum_\psi \big| \langle \psi | nm \rangle \big|^2 |\psi\rangle\langle \psi|,
\end{equation}
the long-time average of the density matrix $|\Psi(t)\rangle\langle\Psi(t)|$. Here, $|\Psi(0)\rangle = |nm\rangle $ and the sum runs over the eigenstates of~\eqref{eq: 2d AA model}. Given the density distribution due to this diagonal ensemble, $\varrho_{nm}(n',m') = \langle n' m'|\rho_{nm}|n'm'\rangle$, we define its participation ratio similarly to \eqref{eq: PR wave function} as
\begin{equation}
    \mathrm{PR}_{nm} = \Big(\sum_{n'm'} \big|\langle n' m'|\rho_{nm}|n'm'\rangle \big|^2\Big)^{-1}.
    \label{eq: diagonal ensemble PR}
\end{equation}
Broadly speaking, this participation ratio captures over how many lattice sites a particle initially localised to a given site will expand. It is plotted in Fig.~\ref{fig: dynamical participation ratio} as a function of the initial site for $\lambda=2$ and 40, along with two representative diagonal ensembles for each. 
At the critical point, diagonal ensembles show fractal properties but altogether appear delocalised on the simulated length scales; indeed, $\mathrm{PR}\simeq L^2$ for most initial sites. 
For $\lambda\gg2$, the crucial difference between low-disorder lines and the rest of the lattice appears very pointedly in the participation ratios. For most initial sites (including the origin), the bulk of the probability distribution remains localised close to the starting point, with exponentially suppressed probability of reaching the network of low-disorder lines. On the contrary, starting from such a line leads to fast delocalisation across the network of low-disorder lines, in a pattern similar to the eigenstate shown in Fig.~\ref{fig: wave function}(c). We demonstrate in Appendix~\ref{app: ballistic} that this expansion occurs with constant speed, i.e., that it is ballistic.

To quantify whether the diagonal ensembles in Fig.~\ref{fig: dynamical participation ratio} are extended, we evaluated the radial total probability $P(r)$ and mean radial density $\overline{\rho(r)}$, defined by
\begin{align}
    P(r) &= \sum_{d=r-1/2}^{r+1/2} \langle n' m'|\rho_{nm}|n'm'\rangle, &
    \overline{\rho(r)} &= \frac{P(r)}{N(r)},
    \label{eq: radial distribution}
\end{align}
where $N(r)$ is the number of sites whose distance $d=\sqrt{(m-m')^2+(n-n')^2}$ from the initial site is between $r-1/2$ and $r+1/2$. 
A constant $P(r)$ indicates states that are uniformly extended along individual lines, while a constant $\overline{\rho(r)}$ corresponds to states extended over the whole plane. 
Indeed, for $\lambda\le 2$, $\overline{\rho(r)}$ tends to a constant${}\approx L^{-2}$ for most initial sites, indicating  almost uniform distributions compatible with extended states. 
For $\lambda\gg2$, $P(r)$ decays exponentially for most initial sites before it is eventually capped by the contribution of delocalised lines: Starting from such a line, by contrast, leads to a nearly constant $P(r)$, consistent with uniform delocalisation in one dimension.
%
%

\section{Generality of effect} 
The model we considered so far is fine-tuned in the sense that the quasiperiodic modulation must be applied at precisely  $\pm45^\circ$ to get infinitely long weakly modulated lines.
%
In this section, we study the effects of moving  away from this ideal case and  of adding random disorder.

We furthermore show that the ground state localisation transition of the 2DAA model belongs to the same universality class as the corresponding continuum quasiperiodic Hamiltonian~\cite{Viebahn2019Matter-WaveLattice}, analogous to what we found earlier in the one-dimensional case~\cite{Szabo2018Non-power-lawQuasicrystals}. While this universality
does not extend directly to all excited states, it does indicate that
the physics of the two models are very closely related to one another.

\begin{figure*}
    \centering
    \includegraphics{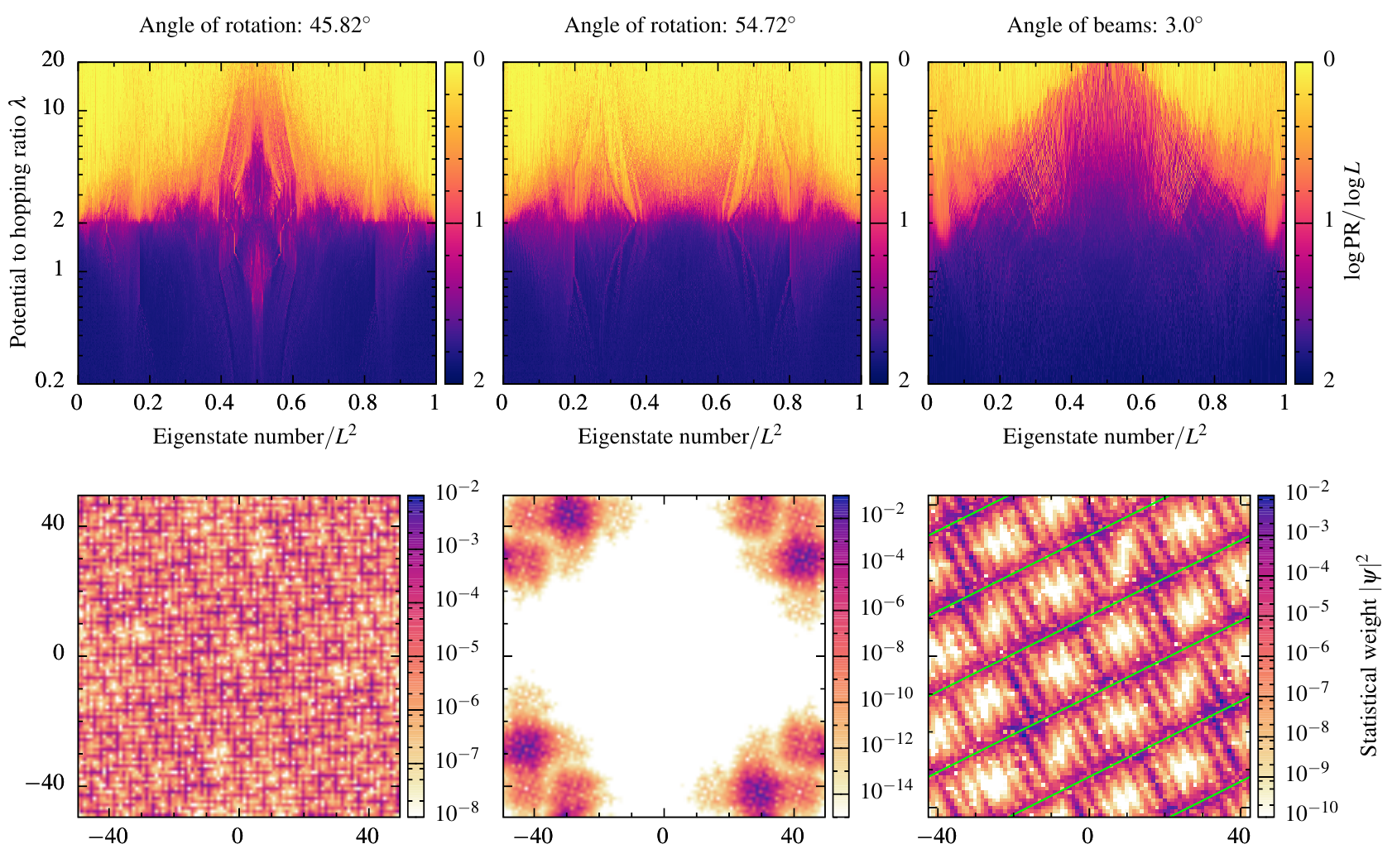}
    \caption{\textbf{Top row:} Participation ratios of all eigenstates of~\eqref{eq:SM: 2d AA model} as a function of $\lambda$ for $\beta_{1,2}=(69,71)/99$ (left column) and $(58,82)/99$ (middle column; $L=99$), and of the non-self-dual model discussed in the text (right column). 
    The pattern of partially extended states seen at $45^\circ$ largely survives for the weakly tilted model, but not at substantial rotations. The non-self-dual model shows a similar localisation pattern, even though the underlying mechanism is different.\\
    \textbf{Bottom row:} Wave function weight $|\psi|^2$ for the eigenstate of the largest PR of each model at $\lambda=5$. The slightly tilted model shows a network of lines reminiscent of the $45^\circ$ case, although the lines are broken up into shorter segments to follow low-disorder areas. At larger tilts, this gives way to exponential localisation. In the non-self-dual model, adding two modulations with similar wave vectors gives rise to wide channels (green lines) along which the wave function is extended, with quasirandom links connecting neighbouring channels. }
    \label{fig:SM: rotated}
\end{figure*}

\subsection{Tilted lattices}
\label{sec:SM: tilted}

To demonstrate the stability of partially extended states against geometric imperfections such as small tilts of the nominally diagonal [$(m+n)$ and $(m-n)$ in~\eqref{eq: 2d AA model}] modulation, we considered self-dual models~\eqref{eq:SM: 2d AA model} where the modulation still takes the form of two perpendicular cosine potentials, but at an angle different from $45^\circ$ to the axes. In particular, we considered the two cases $(\beta_1,\beta_2) = (69,71)/99$ and $(58,82)/99$ with periodic boundary conditions on $99\times99$ sites, corresponding to modulation angles${}\approx45.82^\circ$ and${}\approx54.72^\circ$, respectively. The participation ratios of all eigenstates are plotted in Fig.~\ref{fig:SM: rotated}, along with the respective eigenstates of the highest PR for $\lambda=5$. For modulation angles close to $45^\circ$, the structure of the spectrum and eigenstates remains similar as long as $\lambda$ is not much larger than 2. One can understand these eigenstates as still living on low-disorder lines, which, however, no longer align perfectly with the lattice axes. As long as the modulation is not too strong, it nonetheless remains possible for the wave function to hop between adjacent lattice lines and thus follow the low-disorder region. For larger rotation angles, this structure is effectively destroyed; nevertheless, there is still no sharp transition at $\lambda=2$ for all eigenstates, as also found recently in Ref.~\cite{Huang2019MoireSystems}.

A different kind of partially extended state is generated by related non-self-dual models in which the wave vectors of the two cosine terms in Eq.~\eqref{eq:SM: 2d AA model} are no longer perpendicular but instead make small angles with one another. As an example, we consider in Fig.~\ref{fig:SM: rotated} also a model with wave vectors $2\pi(77,36)/85$ and $2\pi(75,40)/85$ with periodic boundary conditions on $85\times85$ sites. 
We again see many eigenstates with large participation ratios even at large $\lambda$.
These eigenstates, however, have a different structure:
Adding two cosines of similar wave vectors gives rise to wide channels (green lines in Fig.~\ref{fig:SM: rotated}) along the direction of the average wave vector in which the potential remains close to zero. 
Wave functions at $E\approx 0$ can readily delocalise along these channels. Furthermore, there appear to be paths through which different channels can couple to one another, leading to a different kind of partially extended ``mesh''. This construction is manifestly more robust than the self-dual one as long as the two modulating wave vectors make small angles with one another.

\begin{figure*}
    \centering
    \mbox{\includegraphics{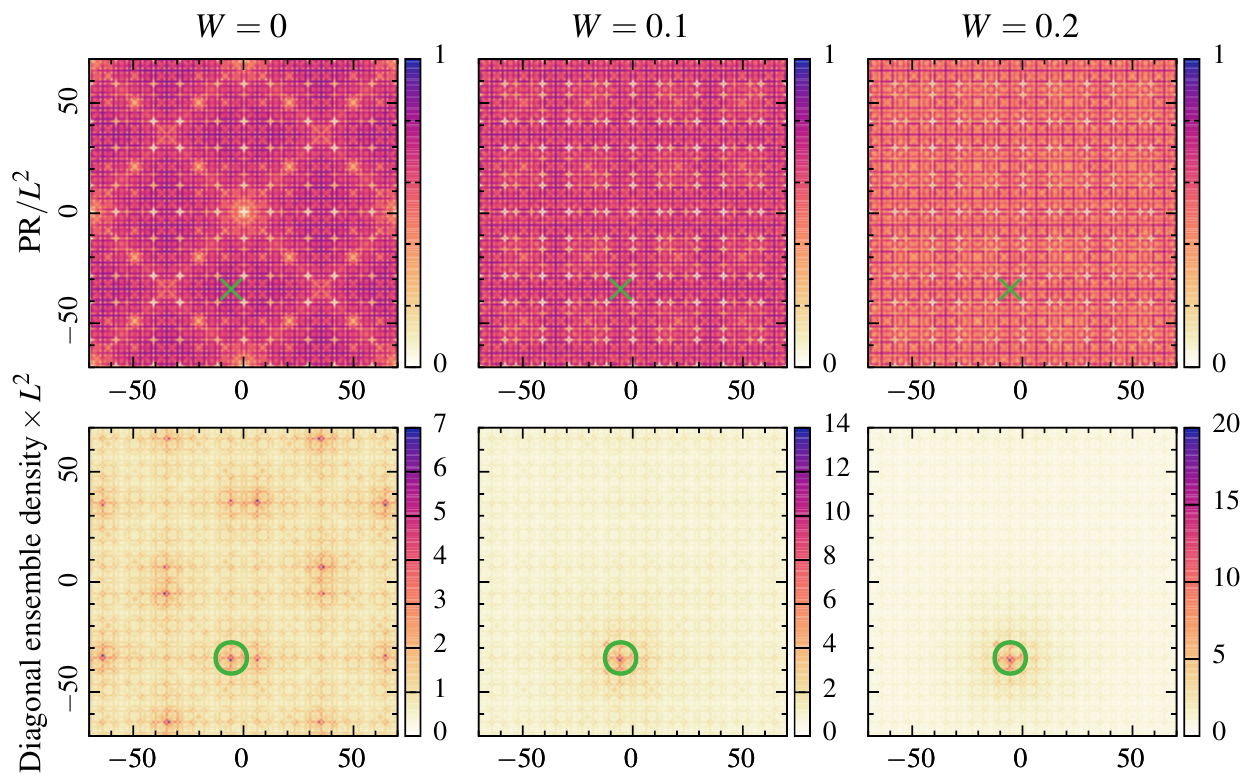}%
    \includegraphics{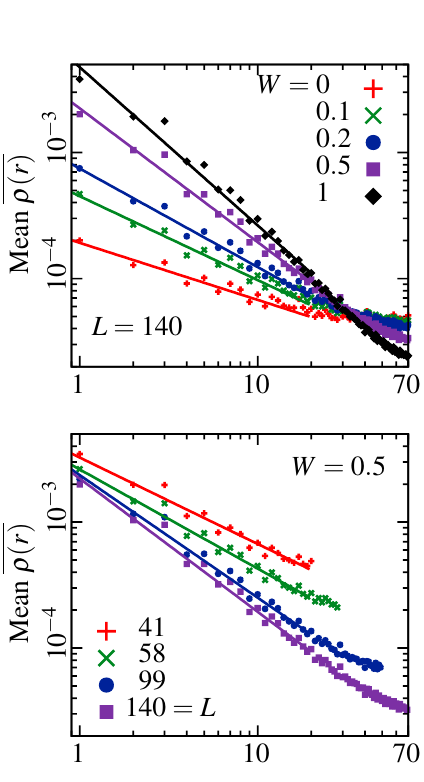}}
    \caption{\textbf{Top left:} Diagonal ensemble participation ratio of the disordered 2DAA model for $\lambda=1.8$, $\beta=99/140$, and three values of disorder strength $W$, averaged over 64 disorder realisations. The typical PR changes little over this region, however, certain features (e.g., diagonal modulation) seen in the absence of disorder disappear. Low-disorder lines, which dominate the dynamics of the clean system at large $\lambda$, are visible as darker lines, indicating their larger than average PR.\\
    \textbf{Bottom left:} Diagonal ensemble density after expanding from site $(-35,-6)$ in the same models. Some large-scale, symmetry-related features of the clean model are lost and most of the statistical weight remains close to the original site. Nevertheless, short-range features remain largely intact and substantial weight can be seen away from the initial site.\\
    \textbf{Right:} Radial density distributions averaged over all initial sites as a function of disorder strength (top) and system size (bottom). For the simulated system sizes, all datasets are dominated by power-law decays. We note that for much larger distances, Anderson localisation due to the random disorder would lead to exponential decay.}
    \label{fig:SM: disorder}
\end{figure*}

\subsection{Random disorder}

Random disorder always leads to exponential localisation in two dimensions, although with exponentially large localisation lengths~\cite{Abrahams1979ScalingDimensions}. This poses the question whether the effects of quasiperiodicity discussed so far 
can ever be observed in a real system, where some random disorder will inevitably be present. In order to study this, 
we have added random on-site disorder to the 2DAA Hamiltonian with $\lambda=1.8$ (i.e., somewhat below the self-dual point), where most eigenstates are still extended. The disorder is drawn from a uniform distribution on $[-W/2,W/2]$ for several values of $W$. The diagonal ensemble participation ratio~\eqref{eq: diagonal ensemble PR} is shown in Fig.~\ref{fig:SM: disorder} at $W/J=0$, 0.1, and 0.2 (averaged over 64 disorder configurations), together with a representative diagonal ensemble for each. In the clean system, particles released from almost all initial sites delocalise almost uniformly over the whole system; furthermore, precursors to the fractal structures characteristic of the self-dual point can be seen. With increasing random disorder, these features become less pronounced and, eventually, much of the long-time density distribution remains close to the initial lattice site. 
However, due to the very large localisation lengths  for weak random disorder in two dimensions, the main effects remain robust for moderate disorder values: While large-scale fractal behaviour gets washed out,  substantial participation ratios remain on experimentally relevant system sizes, even at relatively large disorder strengths on the order of $0.1J$. 
Furthermore, very different behaviours are observed for different initial sites; namely, initial sites on the weak-disorder network observed in the $\lambda\gg 2$ case retain a substantially larger PR than most other lattice sites. 

Even though the diagonal ensemble displays a strong maximum around the initial site, substantial statistical weight can still be seen far away from it even at relatively strong disorder, indicating that it is not exponentially localised. To quantify this, we evaluated the mean radial distribution~\eqref{eq: radial distribution} of diagonal ensemble densities for a range of $W$, averaged over initial sites. These results are plotted in the right hand panels of Fig.~\ref{fig:SM: disorder}: the distribution decays only as a power of the distance from the initial site; interestingly, the exponent of the power law depends on $W$ (top right panel). At large distances comparable to the system size, the distribution tends to a constant, but as shown in the bottom right panel, this is most likely a finite size effect.

To understand these results, it is important to remember that random disorder in two dimensions localises weakly, with an exponentially large localisation length~\cite{Abrahams1979ScalingDimensions}. It is likely that, even at $W=1$, these localisation lengths are far larger than the system sizes available to our simulations and ultracold atom experiments. Therefore, we are confident that the relatively weak random disorder typically caused by experimental imperfections will not hinder the observation of the quasicrystal physics.

\subsection{Universality between lattice and continuum models}

We consider an experimentally inspired \cite{Viebahn2019Matter-WaveLattice} continuum Hamiltonian with eightfold rotational symmetry,
\begin{equation}
    H = \frac{\hat{\mathbf{p}}^2}{2m} + \sum_{n=0}^3 V_n\cos^2(\mathbf{k}_n\cdot\mathbf{r}),
    \label{eq:SM: continuum Hamiltonian}
\end{equation}
where all the $V_n$ are equal and the $\mathbf{k}_n$ have the same magnitude and make $45^\circ$ angles with one another. This model is continuously connected to the self-dual 2DAA model (\ref{eq: 2d AA model}) considered above. Namely, if $V_0=V_2\gtrsim 8E_\mathrm{r}\gg V_1 = V_3$, where $E_\mathrm{r} =\hbar^2 k^2/(2m)$ is the recoil energy, particles will be confined to the minima of the square lattice generated by the strong terms, while the weak terms act as the quasiperiodic modulation of the resulting tight-binding model~\cite{Luschen2018Single-ParticleLattice}. 

We showed earlier~\cite{Szabo2018Non-power-lawQuasicrystals} that the ground state localisation transition of the one-dimensional Aubry--Andr\'e model and an analogous continuum quasiperiodic Hamiltonian belong to the same universality class, even though these classes are not always adequately described by critical exponents. We now show that this is also the case for the 2DAA model and the continuum Hamiltonian~\eqref{eq:SM: continuum Hamiltonian}. 

Periodic approximants of~\eqref{eq:SM: continuum Hamiltonian} for numerical simulation were constructed by replacing the $\mathbf{k}_n$ with
\begin{align*}
    \mathbf{k}_0 &= k(1,0) &\mathbf{k}_2 &= k(0,1) &
    \mathbf{k}_{1,3} &= k \left(\frac{M}L, \pm \frac{M}L\right),
\end{align*}
where $M/L$ is a rational approximation of $1/\sqrt{2}$, leading to a potential that is periodic on a square of size $L\times L$ wavelengths.
The resulting Hamiltonian was represented in the momentum basis $|a,b\rangle = |\mathbf{p}_0 + 2k(a,b)/L\rangle$, and the ground states of both models were obtained to a high accuracy with the L\'anczos method which allows us to attain much larger systems than would be possible with full diagonalisation. 

We have used the ground state curvature~\cite{Lieb2002SuperfluidityGases,Roth2003PhaseSuperlattices,Szabo2018Non-power-lawQuasicrystals}, defined respectively as
\begin{align}
    \Gamma &= \left.\frac1J\frac{E_\Theta-E_0}{(\Theta/L)^2}\right|_{\Theta\to0}, &
    \Gamma &= \left.\frac{\pi^2}{E_\mathrm{r}}\frac{E_\Theta-E_0}{(\Theta/L)^2}\right|_{\Theta\to0}
    \label{eq:SM: ground state curvature}
\end{align}
for lattice and continuum systems, as an order parameter of the extended phase. In~\eqref{eq:SM: ground state curvature}, $E_\Theta$ and $E_0$ are the ground state energies of the periodic system of $L\times L$ sites with periodic boundary conditions twisted by angle $\Theta$ and without twist, respectively~\footnote{For lattice systems, the twist can be implemented by adding a complex phase $\Theta/L$ to all hopping terms. For continuum systems, $\mathbf{p}_0$ is set to $k\Theta/L\pi$.}. In the 2DAA model, this order parameter vanishes at the self-dual point $\lambda=2$, indicating a ground state localisation transition. Likewise, the ground state curvature of the continuum model~\eqref{eq:SM: continuum Hamiltonian} indicates a localisation transition at a finite value of $V$, which is, however, not fixed by any special property of the model, and likely depends on the parameters of the Hamiltonian. 

\begin{figure}
    \centering
    \includegraphics{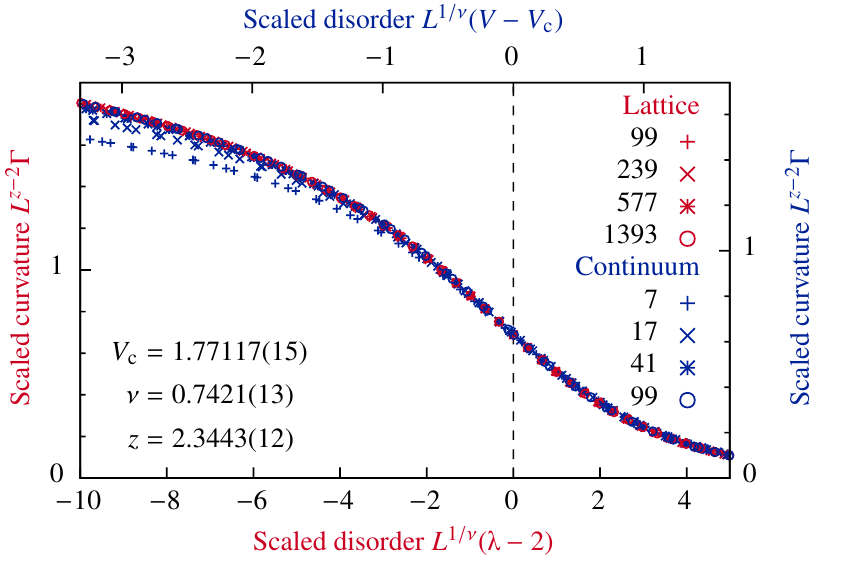}
    \caption{Finite size scaling of the ground state curvature $\Gamma$ for the 2DAA model (red dots) and the continuum model~\eqref{eq:SM: continuum Hamiltonian} (blue dots). Both sets of data share the same critical exponents and fit to the same scaling curve, indicating that their quantum localisation transitions belong to the same universality class. }
    \label{fig:SM: universality}
\end{figure}

To extract the transition point and critical exponents of the transition accurately, we have used finite-size scaling~\cite{fss,Melchert2009} on periodic approximants of different sizes $L$. Fitting results are shown in Fig.~\ref{fig:SM: universality}. The two models share critical exponents and fit well to the same scaling functions, indicating they indeed belong to the same universality class. (Since the continued fraction expansion of $1/\sqrt{2}$ is periodic, it was expected to admit well-defined critical exponents~\cite{Szabo2018Non-power-lawQuasicrystals}.) Although this universality does not extend directly to excited states, it does indicate that the physics of the two models are closely related to one another and that observations we made on the lattice model may apply to the continuum version as well.

\section{Conclusion} 
We studied a self-dual generalisation of the celebrated Aubry--Andr\'e model to two dimensions.
We found that the localisation transition is much richer than in one dimension, with localised and (partially) extended states interspersed in a significant part of the spectrum. This effect, which is present up to large quasiperiodic modulations, stems from the peculiar long-range ordered structure of the quasiperiodic potential and provides a fascinating alternative to the typical notion of rare regions in randomly disordered systems. This complex spectrum with a mixture of ballistically expanding and fully localised states gives rise to intriguing transport properties and provides an interesting new opportunity for studying the fate of many-body localisation in two dimensions. Namely, our model introduces locally one-dimensional ``rare lines'' into an Anderson localised system, which is qualitatively different from both the two-dimensional rare regions in typical disordered systems and the complete absence of rare regions normally envisaged in the quasiperiodic case.

Like in the 1D case~\cite{Szabo2018Non-power-lawQuasicrystals,Agrawal2019UniversalityChains}, we find that the ground-state localisation transitions of different models with the same quasiperiodicity belong to the same universality class. We therefore believe that corresponding continuum models, such as the eightfold symmetric optical quasicrystal studied in Ref.~\cite{Viebahn2019Matter-WaveLattice}, will share some of the physics found here. 
In future works it will be interesting to test, for example via the gap-labelling theorem~\cite{Bellissard1992GapOperators}, whether these models are also topologically equivalent. In general, quasiperiodic systems will enable new studies of the interplay between localisation and topology, as they inherit topological properties from their higher-dimensional parent Hamiltonians~\cite{Kraus2013Four-DimensionalQuasicrystal}. Furthermore, they might help realise novel topological structures in driven systems without the detrimental influence of Floquet heating~\cite{Weinberg2015MultiphotonLattices,Reitter2017InteractionLattice}, as many-body localised states can remain stable in driven systems due to their non-ergodic character~\cite{Bordia2017PeriodicallySystem}.

\section*{Acknowledgements}
We thank Anushya Chandran, Trithep Devakul, David Huse, and Laurent Sanchez-Palencia for stimulating and helpful  discussions. 
This work was partly funded by the European Commission ERC Starting Grant Quasicrystal, EPSRC Grant No. EP/R044627/1, and the EPSRC Programme Grant DesOEQ (EP/P009565/1).

\appendix

\begin{figure*}
    \centering
    \includegraphics{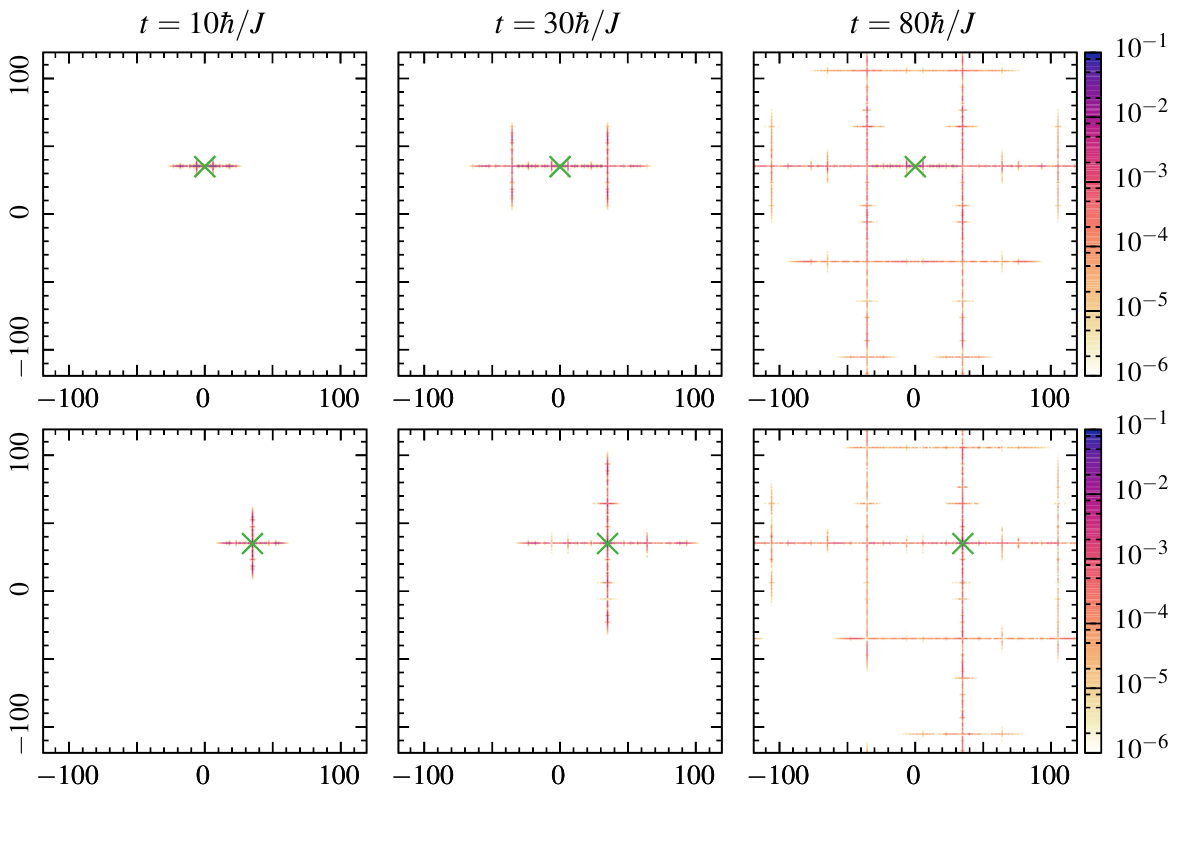}
    \includegraphics{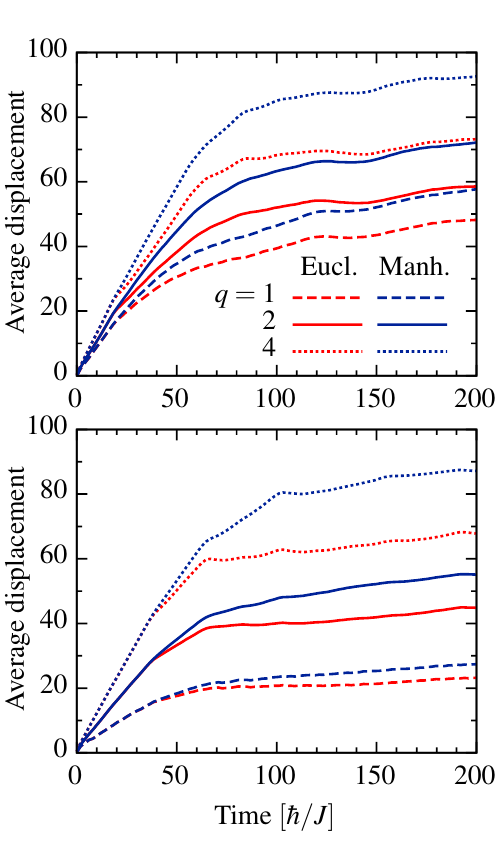}
    \caption{\textbf{Left:} Wave function weight $|\psi|^2$ of an initially localised wave packet at three different times after quenching onto the 2DAA model with $\lambda=40$. The initial sites $(0,35)$ (top row) and $(35,35)$ (bottom row) are marked by the green \textsf{X}, and both lie on the network of delocalised lines discussed in the main text. The wave packet only expands along the same network of low-disorder lines, over a region of size proportional to time, consistent with ballistic expansion.\\
    \textbf{Right:} Average displacement (first, second, and fourth moments of Euclidean and Manhattan distances to the original site) as a function of time. Initially, all measures are linear in time, as expected for ballistic expansion; they saturate at longer times due to the finite simulation box.}
    \label{fig:SM: time evolution}
\end{figure*}

\section{Aubry duality in one and two dimensions}
\label{app: aubryduality}

Under the Fourier transform
\begin{equation}
    b_k = \frac1{\sqrt{\mathcal{N}}} \sum_n \exp\left(2\pi i\beta kn\right) a_n,
    \label{eq:SM: aubryduality}
\end{equation}
the one-dimensional Aubry--Andr\'e model 
\begin{align}
    H &= -J \sum_n \big( a^\dagger_{n}a^{\phantom\dagger}_{n+1} + \mathrm{H.c.}\big) - \lambda J\sum_n \cos(2\pi\beta n) a^\dagger_{n}a^{\phantom\dagger}_{n}
    \label{eq:SM: 1d AA model}
\end{align}
$(\beta\not\in\mathbb{Q})$ can be reexpressed as
\begin{align}
    H = -\frac{\lambda J}2 \sum_k \big( b^\dagger_{k}b^{\phantom\dagger}_{k+1} + \mathrm{H.c.}\big) - 2 J\sum_k \cos(2\pi\beta k) b^\dagger_{k}b^{\phantom\dagger}_{k}.
    \label{eq:SM: 1d AA model dual}
\end{align}
It is easy to see that this Hamiltonian is formally equivalent to \eqref{eq:SM: 1d AA model} with $\lambda$ replaced by $4/\lambda$, and the overall energy scale rescaled by a factor of $\lambda/2$. In particular, \eqref{eq:SM: 1d AA model} and \eqref{eq:SM: 1d AA model dual} are the same for $\lambda=2$, indicating a self-dual point.

Similar self-dual models can also be constructed in two dimensions. Consider the following tight binding Hamiltonian on the square lattice:
\begin{align}
    H &= -J \sum_{nm} \big[a^\dagger_{nm} \big(a^{\phantom\dagger}_{n+1,m} + a^{\phantom\dagger}_{n,m+1}\big) + \mathrm{H.c.}\big] 
    \label{eq:SM: 2d AA model}\\
    &- \lambda J\sum_{nm} \big\{\cos[2\pi (\beta_1 n + \beta_2 m)] 
    + \cos[2\pi (\beta_2 n - \beta_1 m)]\big\} 
    a^\dagger_{nm}a^{\phantom\dagger}_{nm},\nonumber
\end{align} 
where $\beta_{1,2}$ are irrational. In terms of the reciprocal space operators
\begin{equation}
    b_{jk} = \frac1{\sqrt{\mathcal{N}}} \sum_n \exp\Big[2\pi i \big\{ \beta_1 (jn-km) + \beta_2 (jm+kn) \big\}\Big] a_{nm},
    \label{eq:SM: 2d aubryduality}
\end{equation}
Equation~\eqref{eq:SM: 2d AA model} can be rewritten as
\begin{align}
    H &= -\frac{\lambda J}2 \sum_{jk} \big[b^\dagger_{jk} \big(b^{\phantom\dagger}_{j+1,k} + b^{\phantom\dagger}_{j,k+1}\big) + \mathrm{H.c.}\big] 
    \label{eq:SM: 2d AA model dual}\\
    &- 2J\sum_{jk} \big\{\cos[2\pi (\beta_1 j + \beta_2 k)] 
    + \cos[2\pi (\beta_2 j - \beta_1 k)]\big\} 
    b^\dagger_{jk}b^{\phantom\dagger}_{jk},\nonumber
\end{align} 
which is again formally equivalent to \eqref{eq:SM: 2d AA model} with $\lambda\leftrightarrow 4/\lambda$ and scaled by a factor of $\lambda/2$. We note that the 2DAA model~\eqref{eq: 2d AA model} is~\eqref{eq:SM: 2d AA model} with $\beta_1=\beta_2$, while two of the models in Sec.~\ref{sec:SM: tilted} follow by setting $\beta_1\neq\beta_2$.

In principle, it is also possible to choose two arbitrary wave vectors $(\beta_1,\beta_2)$ for the cosine terms in \eqref{eq:SM: 2d AA model} and construct a duality transformation that exchanges the hopping and potential terms. However, unless the two wave vectors are perpendicular to each other, the wave vectors in the dual Hamiltonian will be different from those in the original, and thus the transformation does not give rise to a self-dual point.

We finally point out that the parameters $\beta$ must be irrational in order for the transformation to preserve the full Hilbert space of the tight binding lattice. Indeed, if $\beta_{1,2}$ were rational, $M_{1,2}/L$, $b_{jk} = b_{j+L,k+L}$; that is, at most $L^2$ creation operators would remain independent. In our simulations, we have studied \eqref{eq:SM: 2d AA model} for $\beta_1=\beta_2$ on a tight-binding lattice with $L^2$ sites with periodic boundary conditions, which implies that $\beta_{1,2}$ must be rational with denominator $L$. This is now not a problem as the original Hilbert space consists only of $L^2$ modes; however, one has to be careful to choose $M_{1,2}$ such that none of the reciprocal space modes \eqref{eq:SM: 2d aubryduality} coincide.

\section{Ballistic expansion on low-disorder lines}
\label{app: ballistic}

We calculated the time evolution of particles initially localised on single sites using exact diagonalisation:
\begin{equation}
    |\psi(t)\rangle = \sum_n |n\rangle e^{-iE_nt} \langle n |\psi(0)\rangle,
\end{equation}
where $|n\rangle$ are the eigenstates of the 2DAA Hamiltonian~\eqref{eq: 2d AA model} with energy $E_n$, and $|\psi(0)\rangle$ has support only on site $(x_0,y_0)$. We show the resulting wave functions for $\lambda=0$ with initial sites $(35,0)$ and $(35,35)$ in Fig.~\ref{fig:SM: time evolution}. We note that both initial sites lie on the network of delocalised lines. As a consequence, the wave functions initially expand along this network while avoiding the rest of the lattice. At longer times, they cover the entire network approximately uniformly, as seen for the late-time diagonal ensembles shown in the main text. Wave functions that do not start on or near this network will never reach it and thus remain localised.

We evaluated the $q$th moments of the wave functions $|\psi(t)\rangle$, defined as
\begin{equation}
    \ell_q = \left\langle \psi(t) \middle| \hat{\ell}^q \middle| \psi(t)\right\rangle^{1/q},
\end{equation}
where we used the Euclidean distance $\sqrt{(x-x_0)^2+(y-y_0)^2}$ and the Manhattan distance $|x-x_0|+|y-y_0|$ as distance operators $\hat\ell$. The first, second, and fourth moments corresponding to both metrics and initial sites are shown in Fig.~\ref{fig:SM: time evolution}. At short times, all curves are perfectly linear: This confirms that the expansion is indeed ballistic and has no multicritical features. At late times, all moments saturate as the wave function expands all across the finite simulation box. It is worth pointing out that while moments of the Euclidean distance deviate from linearity as soon as the wave function reaches an intersection of low-disorder lines, Manhattan distances remain approximately linear, indicating that the ballistic expansion is not impeded by spreading out to several lines.

\bibliography{references}

\end{document}